\documentclass[%
 aip,
 jmp,%
 amsmath,amssymb,
preprint,%
]{revtex4-1}

\usepackage{graphicx}
\usepackage{dcolumn}
\usepackage{bm}
\PassOptionsToPackage{table}{xcolor}
\usepackage{stmaryrd}
\usepackage{marvosym}
\usepackage{setspace}
\usepackage{tikz}
\usetikzlibrary{shapes,arrows,shapes.geometric,fit,matrix,backgrounds,plotmarks}
\PassOptionsToPackage{table}{xcolor}
\usepackage{graphicx}
\usepackage{caption}
\usepackage{hyperref}
\usepackage{subcaption}
\usepackage{framed}
\usepackage{fancybox}
\usepackage{colortbl}
\usepackage{color}
\usepackage{pgfpages}
\usepackage{amsmath,mathtools,amssymb}
\usepackage{balance}
\usepackage{lineno}

\usepackage{lineno}
\usepackage{pgfplots}
\usepackage{pgfplotstable}

\begin{document}


\pgfplotstableread[col sep=comma]{./data/r1s1.csv}\obdata
\pgfplotstableread[col sep=comma]{./data/hill.csv}\hill
\pgfplotstableread[col sep=comma]{./data/adair.csv}\adair
\pgfplotstableread[col sep=comma]{./data/dash.csv}\dash

\title{Managing large-scale scientific hypotheses as uncertain and probabilistic data\\ with support for predictive analytics}

\author{Bernardo Gon\c{c}alves}
 \email{bgonc@lncc.br.}
\author{Fabio Porto}%
\email{fporto@lncc.br.}
\affiliation{ 
$^{1)}$Computer Science Dept., National Laboratory for Scientific Computing (LNCC)
}%

 \homepage{http://dexl.lncc.br.}

\date{\today}

\begin{abstract}
%
The sheer scale of high-resolution raw data generated by simulation has 
motivated non-conventional approaches for data exploration referred as `immersive' and `in situ' query processing of the raw simulation data. Another step towards supporting scientific progress is to enable data-driven hypothesis management and predictive analytics out of simulation results. We present a synthesis method and tool for encoding and managing competing hypotheses as uncertain data in a probabilistic database that can be conditioned in the presence of observations.

\end{abstract}


\maketitle
Simulation laboratories provide scientists and engineers with very large, possibly huge datasets that reconstruct phenomena of interest in high resolution. Some examples are the Johns Hopkins Turbulence Databases,\cite{meneveau2007} and the Human Brain Project (HBP) neuroscience simulation datasets.\cite{markram2006} A core motivation for the open delivery of such data is enabling new insights and discoveries through \emph{hypothesis testing against observations}. 

Nonetheless, while the use case for \emph{exploratory analytics} is well understood and many of its challenges have already been coped with so that high-resolution simulation data are increasingly more accessible,\cite{ailamaki2010,ahmad2010} %
only very recently the use case of hypothesis management has been taken into account for \emph{predictive analytics}.\cite{goncalves2014} There is a pressing call for innovative technology to integrate (observed) data and (simulated) theories in a unified framework.\cite{cushing2013} 

Once parametrized access to large-scale simulation data is delivered, tools for connecting hypothesis formulation and testing into the data-driven science pipeline could open promising possibilities for the scientific method at industrial scale. In fact, the point has just been raised by leading neuroscientists in the context of the HBP, who are incisive on the compelling argument that massive simulation databases should be constrained by experimental data in corrective loops to test precise hypotheses.\cite[p. 28]{fregnac2014}
%
%
%

\begin{figure}[t]\footnotesize
\begin{center}
\tikzstyle{rect}=[rectangle,
                                    thick,
                                    minimum size=25pt,
                                    fill=black!8,
                                    draw=black]
\tikzstyle{rectGreen}=[rectangle,
                                    thick,
                                    minimum size=25pt,
                                    fill=green!15,
                                    draw=black]
\tikzstyle{rectBlue}=[rectangle,
                                    thick,
                                    minimum size=25pt,
                                    fill=blue!9,
                                    draw=black]
\tikzstyle{backDisc}=[rectangle,
			       style=dashed,
                                    rounded corners=3pt,
                                    minimum size=82pt,
                                    minimum width=240pt,
                                    draw=black]
\tikzstyle{backJust}=[rectangle,
			       style=dashed,
                                    rounded corners=3pt,
                                    minimum size=82pt,
                                    minimum width=200pt,
                                    draw=black]
\tikzstyle{box}=[rectangle,
                                    fill=none,
                                    draw=none]
\tikzstyle{diamond1}=[diamond,
                                    thick,
                                    minimum size=45pt,
                                    inner sep=0pt,
                                    fill=black!8,
                                    draw=black]
\tikzstyle{cyl2}=[cylinder,
                                    thick,
                                    fill=black!20,
                                    minimum size=30pt,
                                    inner sep=0pt,
                                    draw=black]
\tikzstyle{edge} = [draw,thick,->, ,bend right] 
\begin{tikzpicture}[scale=1.5]
    \node[backDisc] (backDisc) at (-1.55,4) {};
    \node[backJust] (backJust) at (4.2,4) {};
    \node[box] (disc) at (-3.1,4.73) {\textsf{Context of discovery}};
    \node[box] (just) at (5.15,4.73) {\textsf{Context of justification}};
        
    \node[rect] (ph) at (-3.45,3.7) {
    \begin{tabular}{cc}
	$\!$\textsf{Phenomenon}$\!$\vspace{-3pt}\\
	$\!$\textsf{observation}$\!$\\
    \end{tabular}
};    
    \node[rectGreen] (h) at (-1.65,3.7) {
    \begin{tabular}{cc}
	$\!$\textsf{Hypothesis}$\!$\vspace{-3pt}\\
	$\!$\textsf{formulation}$\!$\\
    \end{tabular}
};    
    \node[rect] (cm) at (0.25,3.7) {
    \begin{tabular}{cc}
	$\!$\textsf{Computational}$\!$\vspace{-3pt}\\
	$\!$\textsf{simulation}$\!$\\
    \end{tabular}
};    
    \node[rectBlue] (t) at (2.75,3.7) {
    \begin{tabular}{cc}
	$\!$\textsf{Testing}$\!$\vspace{-3pt}\\
	$\!$\textsf{against data}$\!$\\
    \end{tabular}
};    
    \node[rect] (pub) at (5.75,3.7) {
    \begin{tabular}{cc}
	$\!$\textsf{Publishing}$\!$\vspace{-3pt}\\
	$\!$\textsf{results}$\!$\\
    \end{tabular}
};    

    \node[diamond1,rotate=90] (hvalid) at (4.27,3.7) {\rotatebox[origin=c]{-90}{\textsf{valid?}}};
    \node[box] (yes) at (4.87,3.9) {\textsf{yes}};
    \node[box] (no) at (3.92,4.13) {\textcolor{red}{\textsf{no}}};

    \draw[->] (ph) to (h);
    \draw[->] (h) to (cm);
    \draw[->] (cm) to (t);
    \draw[->] (t) to (hvalid);
    \draw[->] (hvalid) to (pub);
    \draw[edge] (hvalid) to (h);
                
\end{tikzpicture}
\end{center}
\caption{A view of the scientific method life cycle. It highlights hypothesis formulation and a backward transition to reformulation if predictions `disagree' with observations.}
\label{fig:flow-chart}
\end{figure}
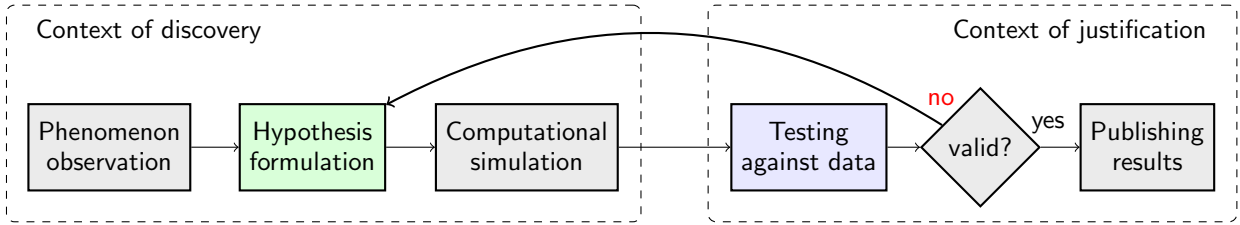

Fig. \ref{fig:flow-chart}$\,$ shows a simplified view of the scientific method life cycle. It distinguishes the phases of exploratory analytics (context of discovery) and predictive analytics (context of justification), and highlights the loop between the hypothesis formulation and testing stages. 
In this article we address the gap currently separating these two stages of the scientific method in the context of data-driven science. 
We present a synthesis method and tool, named $\Upsilon$-DB, for enabling data-driven hypothesis management and predictive analytics in a probabilistic database. It has been demonstrated for simulation data generated from ODE-physiological models,\cite{goncalves2015b} which are available at the Physiome open simulation laboratory.\cite{bassingthwaighte2000} 

\section{Hypothesis Data Management}

Challenges for enabling an efficient access to high-resolution, raw simulation data have been documented from both supercomputing,\cite{meneveau2007} and database research viewpoints;\cite{ailamaki2013} and pointed as key to the use case of exploratory analytics. The extreme scale of the raw data has motivated such non-conventional approaches for data exploration, viz., the `immersive' query processing (move the program to the data),\cite{meneveau2007} or `in situ' query processing in the raw files.\cite{ailamaki2013} 
Both exploit the spatial structure of the data in their indexing schemes.

Simulation data, nonetheless, being generated and tuned from a combination of theoretical and empirical principles, has a distinctive feature to be considered when compared to data generated by high-throughput technology in large-scale scientific experiments such as in astronomy and particle physics surveys. It has a pronounced \emph{uncertainty} component that motivates the use case of hypothesis data management for \emph{predictive analytics}.\cite{goncalves2014} Essential aspects of hypothesis data management can be described in contrast to simulation data management as follows --- Table \ref{tab:hypothesis} summarizes our comparison.

\begin{table}
\caption{Simulation data management vs. hypothesis data management.}
\label{tab:hypothesis}
\begin{spacing}{1.1}
\begingroup\setlength{\fboxsep}{1.5pt}
\colorbox{gray!5}{%
   \begin{tabular}{p{.40\textwidth}|p{.40\textwidth}}
  \rowcolor{gray!15} $\;$\textbf{Simulation data management} & $\;$\textbf{Hypothesis data management}\\
      \hline    
   $\;$Exploratory analytics & $\;$Predictive analytics\\
   $\;$Raw data  & $\;$Sample data\\
   $\;$Extremely large (TB, PB) & $\;$Very large (MB, GB)\\
   $\;$Dimension-centered access pattern & $\;$Claim-centered access pattern\\
   $\;$Denormalized for faster retrieval & $\;$Normalized for uncertainty factors\\
   $\;$Incremental-only data updates & $\;$Probability distribution updates\\  
   \end{tabular}
}\endgroup
\end{spacing}
\end{table}

\begin{itemize}
\item \emph{Sample data}. Hypothesis management shall not deal with the same volume of data as in simulation data management for exploratory analytics, but with samples of it. This is aligned, for example, with the architectural design of CERN's particle-physics experiment and simulation ATLAS, where there are four tier/layers of data. The volume of data significantly decreases from (tier-0) the raw data to (tier-3) the data actually used for analyses such as hypothesis testing.\cite[p. 71-2]{ailamaki2010} 
Samples of raw simulation data are to be selected for comparative studies involving competing hypotheses in the presence of evidence (sample observational data). This principle is also aligned with how data are delivered at model repositories. Since observations are usually less available, only the fragment (sample) of the simulation data that matches in coordinates the (sample) of observations is required out of simulation results for comparative analysis. 
For instance, the graph shown in Fig. \ref{fig:baroreflex} from the Virtual Physiological Rat Project (VPR1001-M) aligns simulation data (heart rates from a baroreflex model) with observations on a Dahl SS rat strain.\cite{beard2010} 
 Here, the simulation is originally set to produce predictions in the time resolution of $0.01$. But since the observational sample is only as fine as $0.1$, the predicted sample is rendered to match the latter in this particular analysis. 

\begin{figure}[t]
\centering
\includegraphics[keepaspectratio,width=.65\textwidth]{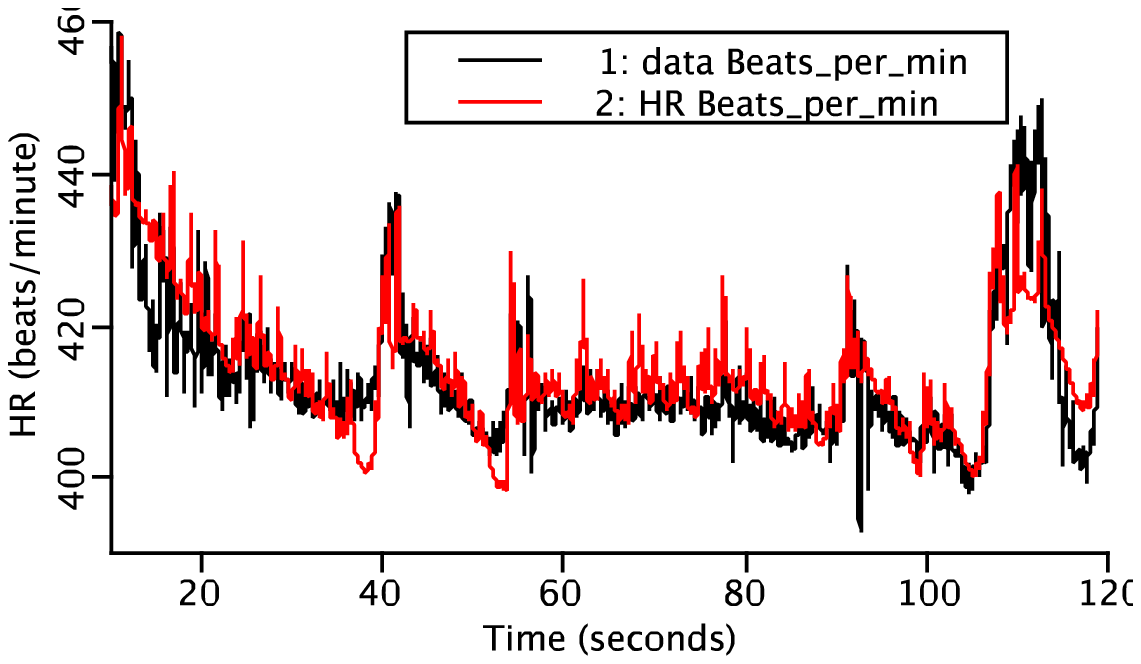}
\caption{Theoretical data generated from a baroreflex model (`HR' in red color) for Dahl SS Rat and its target observations (`data' in black). (source: Bugenhagen et al. \cite{beard2010}).}
\label{fig:baroreflex}
\end{figure}

\item \emph{Claim-centered access pattern}. In simulation data management the access pattern is dimension-centered (e.g., based on selected space-time coordinates) and the data are denormalized for faster retrieval, as typical of Data Warehouses (DW's) and On-Line Analytical Processing (OLAP) applications for decision making --- in contrast to On-Line Transaction Processing (OLTP) applications for daily operations and updates. 
In particular, on account of the so-called `big table' approach, each state of the modeled physical system is recorded in a large, single row of data. This is fairly reasonable for an Extract-Transform-Load (ETL) data ingestion pipeline characterized by \mbox{batch-,} incremental-only updates. Such a setting is in fact fit for exploratory analytics, as entire states of the simulated system shall be accessed at once (e.g., providing data to a visualization system). Altogether, data retrieval is critical and there is no risk of update anomalies. 
Hypothesis management, in contrast, should be centered on claims identified within the hypothesis structure by means of data dependencies. Since the focus is on resolving uncertainty for decision making (which hypothesis is a best fit?), the data must be normalized based on \emph{uncertainty factors}. 
This is key for the correctness of uncertainty modeling and efficiency of probabilistic reasoning, say, in a probabilistic database.\cite[p.30-1]{suciu2011}

\item \emph{Uncertainty modeling}. In uncertain and probabilistic data management,\cite{suciu2011} the uncertainty may come from two sources: \emph{incompleteness} (missing data), and \emph{multiplicity} (inconsistent data). Hypothesis management on sample simulation data is concerned with the multiplicity of prediction records due to competing hypotheses targeted at the same studied phenomenon. Such a multiplicity naturally gives rise to a probability distribution that may be initially uniform and eventually conditioned on observations. Conditioning is an applied \emph{Bayesian inference} problem that translates into database update for transforming the prior probability distribution into a posterior.\cite{goncalves2014}

\end{itemize}

Overall, hypothesis data management is also OLAP-like, yet markedly different from simulation data management. 

\begin{framed}
\noindent
A key point that distinguishes hypothesis management is that a fact or unit of data is defined by its \textbf{predictive content}, not only by its dimension coordinates. Every clear-cut prediction is a claim identified on account of available dependencies. Accordingly, the data should be decomposed and organized for a claim-centered access pattern.
\end{framed}

In what follows we present the use case of hypothesis management and predictive analytics by an example extracted from the Physiome open simulation laboratory (\url{http://www.physiome.org}). 

\section{Use Case: Computational Physiology Hypotheses}

\begin{figure}[t]
\pgfplotsset{every axis legend/.append style={ at={(0.85,0.2)},
anchor=south east}}
\begin{tikzpicture}[y=2cm, x=2cm,font=\sffamily,scale=1.0]
    \begin{axis}[
        height=12cm,
        width=12cm,
        xlabel={\textsf{pO2} [mmHg]},
        ylabel={\textsf{SHbO2} [\%]},
        xtick={0,20, ..., 100}
]
\addplot[color=blue!80!white, style=thick, mark=*, mark size=1.0pt, mark options={fill=blue!80!white}]   table [x=pO2, y=SHbO2.R1s1, col sep=comma] {\obdata};
\addlegendentry{R1s1}
\addplot[color=black, loosely dashed, mark=star, mark size=2.0pt, mark options={fill=black}]  table [x=pO2, y=SHbO2.H, col sep=comma] {\hill};
\addlegendentry{SHbO2.H}
\addplot[color=green!50!gray, solid, mark=square*, mark size=1.0pt, mark options={fill=green!50!gray}]  table [x=pO2, y=SHbO2.D, col sep=comma] {\dash};
\addlegendentry{SHbO2.D}
\addplot[color=red, solid, mark=triangle*, mark size=1.0pt, mark options={fill=red}]  table [x=pO2, y=SHbO2.Ad, col sep=comma] {\adair};
\addlegendentry{SHbO2.Ad}
     \end{axis}
\end{tikzpicture}
\caption{Hemoglobin oxygen saturation hypotheses (SHbO2.\{H, D, Ad\} curves) and their target observations (`R1s1' dataset). (source: Physiome).}
\label{fig:hemoglobin}
\end{figure}
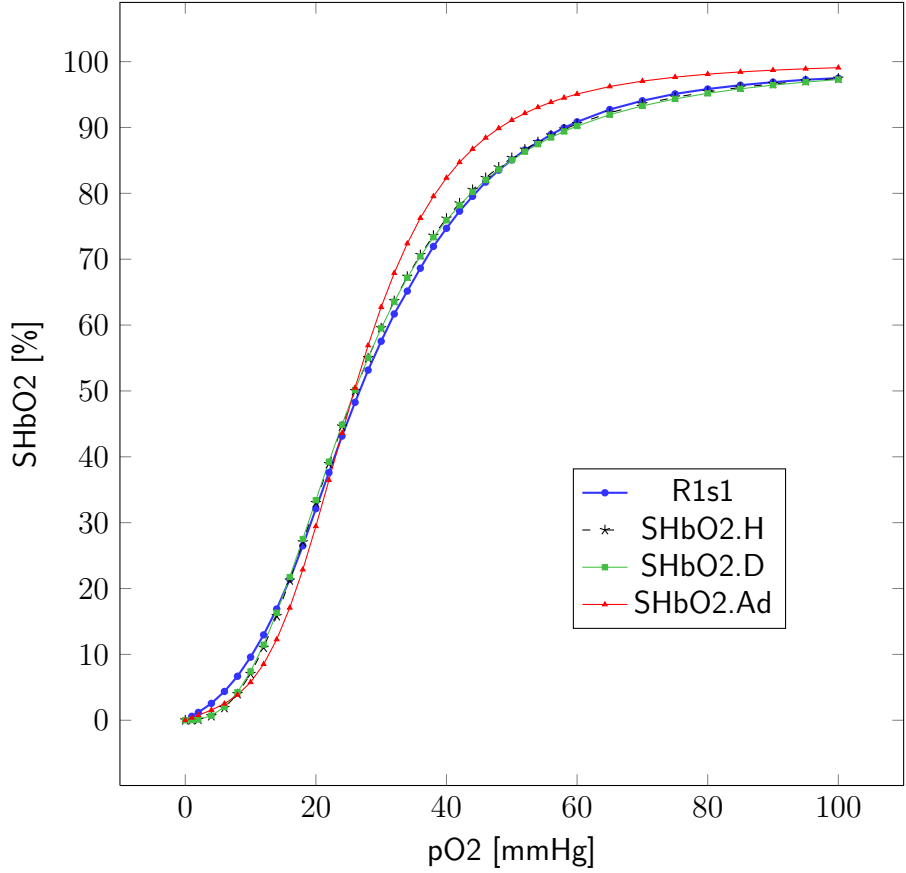

Judy is a researcher in computational physiology who got a set of observations of hemoglobin oxygen saturation from the literature in order to test three different theoretical models stored in an open simulation lab (OSL) against it. She knows about $\Upsilon$-DB, a tool recently plugged into the OSL for data-driven hypothesis management and testing, and decides to try it in view of refining her assessment and reporting, otherwise based on visual analytics only (see Fig. \ref{fig:hemoglobin}). As the models are particularly small, she is able to note by herself that the simplest model is Hill's Equations for $O_2$ binding to hemoglobin (Eqs. \ref{eq:hill}) and then wonders whether it could turn out to be the fittest hypothesis. 
Fig. \ref{fig:case-hemoglobin} shows the textual meta-data easily provided by Judy into the $\Upsilon$-DB system about her study --- special attribute symbols $\phi$ and $\upsilon$ are (resp.) identifiers for the phenomenon and the hypotheses. 

\begin{eqnarray}
\left\{ 
  \begin{array}{lll}
\textsf{SHbO2} &=& \textsf{KO2} \,\cdot\,\textsf{pO2}^\textsf{n}\,/\,(1+\textsf{KO2} \,\cdot\, \textsf{pO2}^\textsf{n})\\
\quad\quad\,\textsf{KO2} &=& \textsf{p50}^{(-\textsf{n})}
\end{array} \right.
\label{eq:hill}
\end{eqnarray}

\begin{figure}[t]
\advance\leftskip-0.2cm
\begin{spacing}{0.9}
\begingroup\setlength{\fboxsep}{4pt}
\colorbox{yellow!15}{%
   \begin{tabular}{c|>{\columncolor[gray]{0.92}}c||l|p{0.49\linewidth}}
  \textsf{HYPOTHESIS}$\;$ & $\upsilon$ & $\;$\textsf{Name} & $\;$\textsf{Description}\\
      \hline    
   & $\;28\;$ & $\;$\textsf{HbO.Hill}$\;$ & Hill's Equation for O2 binding to hemoglobin.\vspace{6pt}\\
   & $\;31\;$ & $\;$\textsf{HbO.Adair}$\;$ & Hemoglobin O2 saturation curve using Adair's 4-site equation.\vspace{6pt}\\
   & $\;32\;$ & $\;$\textsf{HbO.Dash}$\;$ & Hemoglobin O2 saturation curve at varied levels of PCO2 and pH.\vspace{6pt}\\
   \end{tabular}
}\endgroup
\vspace{5pt}
\begingroup\setlength{\fboxsep}{3pt}
\colorbox{blue!4}{%
   \begin{tabular}{c|>{\columncolor[gray]{0.92}}c||p{0.51\linewidth}}
  \textsf{PHENOMENON}$\;$ & $\;\phi\;$ & $\;$\textsf{Description}\\
      \hline    
   & $\;1\;$ & Hemoglobin oxygen saturation with observational dataset from Sevenringhaus 1979.\vspace{6pt}\\ 
   \end{tabular}
}\endgroup
\end{spacing}
\caption{Description of Judy's hypotheses with their original id's from Physiome's OSL.}
\label{fig:case-hemoglobin}
\end{figure}

\section{Hypothesis Encoding}

Scientific hypotheses are tested by way of their predictions. 
In the form of mathematical equations like Eqs. \ref{eq:hill}, hypotheses symmetrically relate aspects of the studied phenomenon. For computing predictions, however, hypotheses are used asymmetrically like \emph{functions}.\cite{simon1966} They take a given valuation over input variables (parameters) to produce values of output variables (the predictions). Interestingly, such an asymmetry can be detected automatically to establish functional dependencies that unravel the structure of the predictive data.\cite{goncalves2015c} 
 
We abstract a system of mathematical equations into a structural equation model $\mathcal S(\mathcal E, \mathcal V)$,\cite{simon1966} where $\mathcal E$ is the set of equations and $\mathcal V$ the set of all variables appearing in them. For instance, note that Hill's Eqs. (\ref{eq:hill}) are $|\mathcal E|=2$, with $\mathcal V=\{\, \textsf{SHbO2},\, \textsf{KO2},\, \textsf{pO2},\, \textsf{n},\, \textsf{p50} \,\}$. Intuitively they do not form a valid (computational model) structure, as $|\mathcal E| \neq |\mathcal V|$. They must be completed by setting domain and parameter values. In this specific case, domain function $f_3(\textsf{pO2})$ and constant functions $f_4(\textsf{n}),\, f_5(\textsf{p50})$ are included into Hill's structure $\mathcal S$.

In view of uncertainty modeling, we need to derive a set $\Sigma_{28}$ of functional dependencies (as `causal' orientations)\cite{simon1966} from Hill's structure. 
We focus on its implicit data dependencies and get rid of constants and possibly complex mathematical constructs. By exploiting Hill's global structure $\mathcal S$, equation $\textsf{KO2} = \textsf{p50}^{(-\textsf{n})}$, e.g., is oriented towards $\textsf{KO2}$, which is then a (prediction) variable \emph{functionally dependent} on (parameters) $\textsf{p50}$ and $\textsf{n}$. Yet a dependency like $\{\textsf{p50},\, \textsf{n}\} \to \{\textsf{KO2}\}$ may hold for infinitely many equations (think of, say, how many polynomials satisfy that dependency `signature'). In fact, we need a way to identify the equation's mathematical formulation precisely, i.e., an abstraction of its data-level semantics. This is achieved by introducing hypothesis id $\upsilon$ as a special attribute in the functional dependency (see $\Sigma_{28}$ below; compare it with Eqs. \ref{eq:hill}).\cite{goncalves2014,goncalves2015c} We adopt here the usual notation for functional dependencies from the database literature, without braces. 

\vspace{-20pt}
\begin{eqnarray*}
\Sigma_{28} = \{\quad \textsf{KO2\;\;n\;\;pO2} \;\upsilon \;\;&\to&\;\; \textsf{SHbO2},\\
\textsf{n\;\;p50} \;\upsilon \;\;&\to&\;\; \textsf{KO2},\\
\phi \;\;&\to&\;\; \textsf{n},\\
\phi \;\;&\to&\;\; \textsf{p50} \quad\}.
\end{eqnarray*}
\vspace{-20pt}

All dependencies containing $\upsilon$ in their left-hand sides mean that the right-hand side variable has its values predicted on account of the values of the variables on the left. The other special attribute, the phenomenon id $\phi$, appears in dependencies of the form $\,\phi \to\, x$. These are informative that $x$ has been identified a parameter whose value is set empirically. That is, it is set `outside' of the hypothesis model, contributing to the parameter valuation that defines a particular trial on the hypothesis and then grounds it for a target phenomenon. 

That is a data representation of the structure of a scientific hypothesis.\cite{goncalves2014} 
%
%
The causal reasoning on the global structure $\mathcal S$ is a challenging, yet accessible problem. It can be performed by an efficient algorithm.\cite{goncalves2015c} As a result, a total `causal' mapping (a bijection) is rendered from set $\mathcal E$ of equations to set $\mathcal V$ of variables, i.e., every equation is oriented towards exactly one of its variables. This technique is provably very efficient, viz., $O(\sqrt{|\mathcal E|}\,|\mathcal S|)$, where $|\mathcal S|$ is the total number of variable appearances in all equations, i.e., a measure of how dense the hypothesis is. So far we have tested it in scale for processing hypotheses whose structures are sized up to $|\mathcal S| \lesssim 1M$, with $|\mathcal E|=|\mathcal V| \leq 2.5\,K$.\cite{goncalves2015c} 

Our technique for hypothesis encoding relies on the availability of the hypothesis structure (its equations) in a machine-readable format such as W3C's MathML. 
In fact, it has been a successful design decision of Physiome's OSL to standardize the MathML-compliant Mathematical Modeling Language (MML) for model specification and sharing towards reproducibility (cf. \url{http://www.physiome.org/jsim/}). The $\Upsilon$-DB system is then provided with an XML wrapper component to extract Physiome's models encoded in MML and infer its causal dependencies. The same can be done for every OSL if its computational models are structured in declarative form such as in a MathML file.

\section{Synthesis Pipeline}

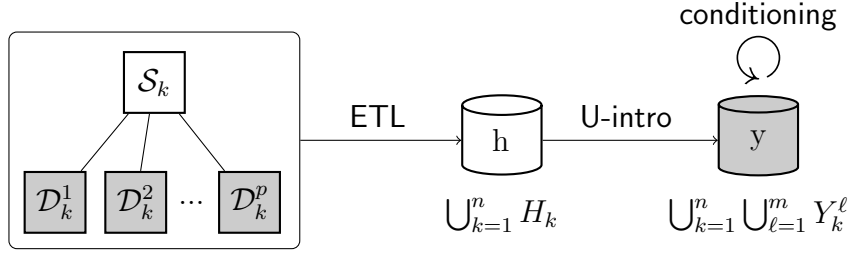
\begin{figure}[t]
\begin{center}
\tikzstyle{rect1}=[rectangle,
                                    thick,
                                    minimum size=23pt,
                                    draw=black]
\tikzstyle{rect2}=[rectangle,
                                    thick,
                                    minimum size=23pt,
                                    fill=black!20,
                                    draw=black]
\tikzstyle{rect3}=[rectangle,
                                    rounded corners=3pt,
                                    minimum size=82pt,
                                    minimum width=110pt,
                                    draw=black]
\tikzstyle{box}=[rectangle,
                                    fill=none,
                                    draw=none]
\tikzstyle{cyl1}=[cylinder,
                                    thick,
                                    minimum size=30pt,
                                    inner sep=0pt,
                                    fill=none,
                                    draw=black]
\tikzstyle{cyl2}=[cylinder,
                                    thick,
                                    fill=black!20,
                                    minimum size=30pt,
                                    inner sep=0pt,
                                    draw=black]
\tikzstyle{edge} = [draw,thick,->,bend left]
\begin{tikzpicture}[scale=1.1]
    \node[rect3] (back) at (0,3) {};
    \node[rect1] (s) at (0,3.7) {$\mathcal{S}_k$};
    \node[rect2] (d1) at (-1.2,2.25) {$\mathcal{D}_k^1$};
    \node[rect2] (d2) at (-0.225,2.25) {$\mathcal{D}_k^2$};
    \node[box] (dot) at (0.45,2.25) {...};
    \node[rect2] (dn) at (1.15,2.25) {$\mathcal{D}_k^p$};
    \node[cyl1,rotate=90] (h) at (4.2,3) {\rotatebox[origin=c]{-90}{h}};
    \node[box] (hlabel) at (4.2,2.1) {$\bigcup_{k=1}^{n}H_k$};
    \node[cyl2,rotate=90] (y) at (7.3,3) {\rotatebox[origin=c]{-90}{y}};
    \node[box] (ylabel) at (7.3,2.1) {$\bigcup_{k=1}^{n}\bigcup_{\ell=1}^{m}Y_k^\ell$};
    \node[box] (cond) at (7.3,4) {\huge\rotatebox[origin=c]{180}{$\circlearrowleft$}};
    \draw[-] (s) to (d1);
    \draw[-] (s) to (d2);
    \draw[-] (s) to (dn);
    \draw[->] (back) to (h);
    \node[box] (etl) at (2.7,3.3) {\textsf{ETL}};    
    \draw[->] (h) to (y);
    \node[box] (etl) at (5.7,3.3) {\textsf{U-intro}};
    \node[box] (etl) at (7.3,4.5) {\textsf{conditioning}};
\end{tikzpicture}
\caption{Synthesis pipeline for processing hypotheses as uncertain and probabilistic data. For each hypothesis $k$, its structure $\mathcal S_k$ is given in a machine-readable format, and all its sample simulation data trials $\bigcup_{i=1}^p\mathcal{D}_k^i$ are indicated their target phenomenon id $\phi$ and loaded into a `big table' $H_k$. Then the synthesis comes into play to read a base of possibly very many hypotheses $\bigcup_{k=1}^n H_k$ and transform them into a probabilistic database where each hypothesis is decomposed into claim tables $\bigcup_{\ell=1}^{m}Y_k^\ell$. A probability distribution is computed for each phenomenon $\phi$, covering all the hypotheses and their trials targeted at $\phi$. This distribution is then updated into a posterior in the presence of observational data.} 
\label{fig:pipeline}
\end{center}
\end{figure}

When Judy inserts a hypothesis $k$, into $\Upsilon$-DB, the system extracts the set of (causal) functional dependencies from its mathematical structure and then she can upload its sample simulation data into a `big table' $H_k$ containing all its variables as relational attributes in the table. For the upload, she chooses a phenomenon to be targeted by the hypothesis simulation data. Both the hypothesis structure and its data are input to the synthesis pipeline shown in Fig. \ref{fig:pipeline}.

Normalization of the `big table' (see Fig. \ref{fig:big-table}), as discussed above, is not desirable because its data are not be updated (only re-inserted if necessary). For hypothesis management, however, the uncertainty has to be decomposed/normalized so that the uncertainty of one claim may not be undesirably mixed with the uncertainty of another claim. In fact, we perform further reasoning, viz., acyclic pseudo-transitive reasoning over functional dependencies,\cite{goncalves2015c} to process $\Sigma_{28}$ into another set $\Sigma_{28}^\prime$ of dependencies. This is ensured to have, for each predictive variable, exactly one dependency with it on the right-hand side and all its uncertainty factors (so-called `first causes') on the left-hand side.\cite{goncalves2015c} 

\vspace{-20pt}
\begin{eqnarray*}
\Sigma_{28}^\prime = \{\qquad\qquad \textsf{n\;\;p50} \;\phi\;\upsilon \;\;&\to&\;\; \textsf{KO2},\\
\textsf{n\;\;p50\;\;pO2} \;\phi\;\upsilon \;\;&\to&\;\; \textsf{SHbO2} \quad\}\\
\end{eqnarray*}

\begin{spacing}{1.0}
\begin{figure}[h]
\centering
\begingroup\setlength{\fboxsep}{3pt}
\colorbox{blue!5}{%
   \begin{tabular}{c|c|c|c||c|c|c|c}
  $H_{28}$ & $\phi$ & $\upsilon$ & $\,$\textsf{pO2}$\,$ & \textsf{KO2} & \textsf{SHbO2} & \textsf{n} & \textsf{p50}\\
      \hline
   \cline{2-8}   
  & $\,$1$\,$ & $\,$28$\,$ & 0 & $\,$1.51207022127057\text{E-4}$\,$ & 0 & $\,$2.7$\,$ & 26\\ 
  & $\,$1$\,$ & $\,$28$\,$ & 0.1 & $\,$1.51207022127057\text{E-4}$\,$ & $\,$3.01697581987324\text{E-7}$\,$ & $\,$2.7$\,$ & 26\\
  & $\,$1$\,$ & $\,$28$\,$ & 0.2 & $\,$1.51207022127057\text{E-4}$\,$ & $\,$1.96043341970514\text{E-6}$\,$ & $\,$2.7$\,$ & 26\\
  & $\,$1$\,$ & $\,$28$\,$ & ... & $\,$1.51207022127057\text{E-4}$\,$ & $\,$...$\,$ & $\,$2.7$\,$ & 26\\
  & $\,$1$\,$ & $\,$28$\,$ & 100 & $\,$1.51207022127057\text{E-4}$\,$ & $\,$9.74346796798538\text{E-1}$\,$ & $\,$2.7$\,$ & 26\\
   \end{tabular}
}\endgroup
\caption{`Big table' $H_{28}$ of hypothesis $\upsilon=28$ (Hill's equation) loaded with one sample trial dataset targeted at phenomenon $\phi=1$. The main predictive variable is hemoglobin oxygen saturation \textsf{SHbO2}, whose values evolve with the values of dimension \textsf{pO2}.}
\label{fig:big-table}
\end{figure}
\end{spacing}

\section{Probabilistic Database}

It is a basic design principle for uncertainty modeling to define exactly one random variable for each actual uncertainty factor (\emph{u-factor}, for short). 
The hypothesis model is itself a theoretical u-factor, whose uncertainty comes from the multiplicity of models targeted at explaining the same phenomenon. Additionally, the multiplicity of trials on each hypothesis (alternative parameter settings) targeted at the same phenomenon gives rise to empirical u-factors. In fact, a hypothesis model can only approximate a phenomenon very well if it is properly tuned (calibrated) for the latter. The probability distribution on a phenomenon must take into account both kinds of u-factors to support hypothesis testing in this broad sense. Each u-factor is captured into a random variable in the probabilistic database.


We carry out the algorithmic transformation from each hypothesis `big table' to the probabilistic database through the probabilistic world-set algebra of the U-relational model, an extension of relational algebra for managing uncertain and probabilistic data.\cite{koch2009,suciu2011}

U-relations have in their schema a set of pairs $(V_i, D_i)$ of \emph{condition columns} to map each discrete random variable $x_i$ 
to one of its possible values (e.g., $x_0 \mapsto 1$). The `world table' $W$, inspired in pc-tables,\cite{suciu2011} 
stores their marginal probabilities. 
Fig. \ref{fig:u-relations} shows the probabilistic U-relational tables synthesized for hypothesis $\upsilon=28$. 
Any row of, say, table $Y_{28}^4$, has the same joint probability distribution $\textsf{Pr}(\theta) \!\approx\! .33$, which is associated with possible world $\theta = \{\,x_0 \!\mapsto\! 1,\, x_1 \!\mapsto\! 1,$ $x_2 \!\mapsto\! 1 \,\}$. The probabilistic inference is performed in aggregate queries by the \textsf{conf} operator based on the marginal probabilities stored in world table $W\!$.\cite{koch2009} 


Such a probabilistic database should bear desirable design-theoretic properties for uncertainty modeling and probabilistic reasoning.\cite[p. 30-1]{suciu2011} In fact, it is in Boyce-Codd normal form w.r.t. the (causal) functional dependencies and its uncertainty decomposition (into marginal probabilities) is recoverable by a lossless join (joint probability distribution).\cite{goncalves2015c}

\begin{figure}[t]
\centering
\begingroup\setlength{\fboxsep}{2pt}
\colorbox{yellow!15}{%
   \begin{tabular}{c|>{\columncolor[gray]{0.92}}c|c|c}
  $\,Y_0\,$ & $V \mapsto D$ & $\phi$ & $\upsilon\!$\\
      \hline    
   & $\,x_0 \mapsto 1\,$ & $\,1\,$ & $\,28\!$\\
   & $\,x_0 \mapsto 2\,$ & $\,1\,$ & $\,31\!$\\
   & $\,x_0 \mapsto 3\,$ & $\,1\,$ & $\,32\!$\\
   \end{tabular}
}\endgroup
\hspace{4pt}
\begingroup\setlength{\fboxsep}{2pt}
\colorbox{yellow!10}{%
   \begin{tabular}{c|>{\columncolor[gray]{0.92}}c||c|c}
  $\,Y_{28}^1\,$ & $\,V \!\mapsto\! D\,$ & $\,\phi\,$ & $\,$\textsf{n}$\,$\\
      \hline    
   & $\,x_1 \!\mapsto\! 1\,$ & $\,1\,$ & $\,2.7\!$\\
   \end{tabular}
}\endgroup
\hspace{4pt}
\begingroup\setlength{\fboxsep}{2pt}
\colorbox{yellow!10}{%
   \begin{tabular}{c|>{\columncolor[gray]{0.92}}c||c|c}
  $\,Y_{28}^2\,$ & $\,V \!\mapsto\! D\,$ & $\,\phi\,$ & $\,$\textsf{p50}$\,$\\
      \hline    
   & $\,x_2 \!\mapsto\! 1\,$ & $\,1\,$ & $\,26$\\
   \end{tabular}
}\endgroup
\vspace{5pt}\\
\begingroup\setlength{\fboxsep}{2pt}
\colorbox{yellow!10}{%
   \begin{tabular}{c|>{\columncolor[gray]{0.92}}c|>{\columncolor[gray]{0.92}}c|>{\columncolor[gray]{0.92}}c||c|c|c}
  $\,Y_{28}^3\,$ & $\,V_0 \!\mapsto\! D_0\,$ & $\,V_1 \!\mapsto\! D_1\,$ & $\,V_2 \!\mapsto\! D_2\,$ & $\,\phi\,$ & $\,\upsilon\,$ & $\,$\textsf{KO2}$\,$\\
      \hline    
   & $\,x_0 \!\mapsto\! 1\,$ & $\,x_1 \!\mapsto\! 1\,$ & $\,x_2 \!\mapsto\! 1\,$ & $\,1\,$ & $\,28\,$ & $\;1.51207022127057\text{E-4}$\\
   \end{tabular}
}\endgroup
\vspace{5pt}\\
\begingroup\setlength{\fboxsep}{3pt}
\colorbox{yellow!10}{%
   \begin{tabular}{c|>{\columncolor[gray]{0.92}}c|>{\columncolor[gray]{0.92}}c|>{\columncolor[gray]{0.92}}c||c|c|c|c}
  $\,Y_{28}^4\,$ & $\,V_0 \!\mapsto\! D_0\,$ & $\,V_1 \!\mapsto\! D_1\,$ & $\,V_2 \!\mapsto\! D_2\,$ & $\,\phi\,$ & $\,\upsilon\,$ & $\,$\textsf{pO2}$\,$ & $\,$\textsf{SHbO2}$\!\!$\\
      \hline    
   & $\,x_0 \mapsto 1\,$ & $\,x_1 \mapsto 1\,$ & $\,x_2 \mapsto 1\,$ & $\,1\,$ & $\,28\,$ & $\,0\,$ & $\,0\,$\\
   & $\,x_0 \mapsto 1\,$ & $\,x_1 \mapsto 1\,$ & $\,x_2 \mapsto 1\,$ & $\,1\,$ & $\,28\,$ & $\,0.1\,$ & $\,$3.01697581987324\text{E-7}\\
   & $\,x_0 \mapsto 1\,$ & $\,x_1 \mapsto 1\,$ & $\,x_2 \mapsto 1\,$ & $\,1\,$ & $\,28\,$ & $\,0.2\,$ & $\,$1.96043341970514\text{E-6}\\
   & $\,x_0 \mapsto 1\,$ & $\,x_1 \mapsto 1\,$ & $\,x_2 \mapsto 1\,$ & $\,1\,$ & $\,28\,$ & $\,...\,$ & $\,...\,$\\
   & $\,x_0 \mapsto 1\,$ & $\,x_1 \mapsto 1\,$ & $\,x_2 \mapsto 1\,$ & $\,1\,$ & $\,28\,$ & $\,100\,$ & $\,$9.74346796798538\text{E-1}\\
   \end{tabular}
}\endgroup
\hspace{4pt}
\begingroup\setlength{\fboxsep}{2pt}
\colorbox{yellow!15}{%
   \begin{tabular}{c|>{\columncolor[gray]{0.92}}c|c}
  $W$ & $V \mapsto D$ & \textsf{Pr}$\!\!\!$\\
      \hline    
   & $\,x_0 \mapsto 1\,$ & $.33$\\
   & $\,x_0 \mapsto 2\,$ & $.33$\\
   & $\,x_0 \mapsto 3\,$ & $.33$\\
   \cline{2-3}
   & $\,x_1 \!\mapsto\! 1\,$ & $\,1\,$\\
   \cline{2-3}
   & $\,x_2 \!\mapsto\! 1\,$ & $\,1\,$\\
   \end{tabular}
}\endgroup
\caption{U-relations synthesized for hypothesis $\upsilon=28$. The model competition on phenomenon $\phi=1$ is captured into the probability distribution associated with random variable $x_0$, see U-relation $Y_0$. The `world table' $W$ stores the marginal probabilities on the random variables. Observe that there is no parameter uncertainty (multiplicity) in this hypothesis $\upsilon=28$, where random variables $x_1$ and $x_2$ are associated with its parameters \textsf{n} and \textsf{p50}. Values of predictive variables are then annotated with the uncertainty coming from their u-factors, which are then combined into a joint probability distribution.}
\label{fig:u-relations}
\end{figure}

\section{Predictive Analytics}

Noticeably, the prior probability distribution on `explanation' random variable $x_0$ is uniform (see world table $W$ in Fig. \ref{fig:u-relations}). Now that Judy's hemoglobin oxygen saturation hypotheses are encoded with their sample simulation data properly stored in the probabilistic database, she is keen to see the results, the hypothesis rating/ranking based on the observed data. The insertion of the latter into $\Upsilon$-DB is straightforward. It is loaded into a relational table (not shown) from a CSV file and associated with phenomenon $\phi=1$.

The system then enables her to carry out Bayesian inference steps that update at each step the prior distribution of her interest to a posterior. In such computational science use cases, as we have \emph{discrete} random variables mapped to the possible values of (numerical) prediction variables whose domain are \emph{continuous} (double precision), the Bayesian inference is applied for normal mean (likelihood function) with a discrete prior (probability distribution).\cite{bolstad2007}

The procedure uses normal density function (Eq. \ref{eq:density}) with standard deviation \mbox{$\sigma$} to compute the likelihood $f(y \,|\, \mu_k)$ for each competing prediction $\mu_k$ given observation $y$. But as we actually have 
a sample of independent observed values $y_1,\,...,\,y_n$ (viz., measured hemoglobin oxygen saturations \textsf{SHbO2} over different oxygen partial pressures \textsf{pO2}). Then, the likelihood $f(y_1,\,...,\,y_n \,|\, \mu_k)$ for each competing trial $\mu_k$, is computed as a product $\textstyle\prod_{j=1}^n f(y_j \,|\, \mu_{kj})$ of the single likelihoods $f(y_j \,|\, \mu_{kj})$.\cite{bolstad2007} Bayes' rule is then settled by (Eq. \ref{eq:sample-bayes}) to compute the posterior $p(\mu_k \,|\, y_1,\,...,\,y_n)$ given prior $p(\mu_k)$. 

\vspace{-25pt}
\begin{eqnarray}
f(y \,|\, \mu_k) \!\!&=&\!\! \frac{1}{\sqrt{2 \pi \sigma^2}}\, e^{-\frac{1}{2\sigma^2}(y-\mu_k)^2}\label{eq:density}
\end{eqnarray}
\begin{eqnarray}
p(\mu_k \,|\, y_1,\,\hdots,\,y_n) \!&=&\! \frac{\textstyle\prod_{j=1}^n f(y_j \,|\, \mu_{kj})\;p(\mu_k)}{\displaystyle\sum_{i=1}^m \displaystyle\prod_{j=1}^n f(y_j \,|\, \mu_{ij})\;p(\mu_i)}\label{eq:sample-bayes}
\end{eqnarray}

\begin{spacing}{1.0}
\begin{figure}[t]
\centering
\begingroup\setlength{\fboxsep}{3pt}
\colorbox{yellow!15}{%
   \begin{tabular}{c|>{\columncolor[gray]{0.92}}c|>{\columncolor[gray]{0.92}}c||c|c|c|c}
  \textsf{STUDY} & $\phi$ & $\upsilon$ & \textsf{pO2} & \textsf{SHbO2} & \textsf{Prior} & \textsf{Posterior}\\
      \hline    
   & $1$ & $32$ & $1$ & $\,0.178973375779681\text{E-3}\,$ & $.333$ & \cellcolor{gray!25} $.335441$\\
   & $1$ & $28$ & $1$ & $\,0.151184162020125\text{E-3}\,$ & $.333$ & $.335398$\\
   & $1$ & $31$ & $1$ & $\,3.789100566457180\text{E-3}\,$ & $.333$ & $.329161$\\
   \cline{2-7}   
   & $\cdots$ & $\cdots$ & $\cdots$ & $\cdots$ & $\cdots$\\   
   \cline{2-7}   
   & $1$ &  $32$ & $100$ & $\,9.72764121981342\text{E-1}\,$ & $.333$ & \cellcolor{gray!25} $.335441$\\
   & $1$ &  $28$ & $100$ & $\,9.74346796798538\text{E-1}\,$ & $.333$ & $.335398$\\
   & $1$ &  $31$ & $100$ & $\,9.90781330988763\text{E-1}\,$ & $.333$ & $.329161$\\
   \end{tabular}
}\endgroup
\vspace{3pt}
\caption{Results of Judy's analytics on the computational physiology hypotheses. The predictions from each hypothesis are aligned with the observational dataset which is of smaller resolution. The hypothesis rating/ranking is derived from Bayesian inference.}
\label{fig:case-hemoglobin-analytics}
\vspace{-6pt}
\end{figure}
\end{spacing}
%
%

Fig. \ref{fig:case-hemoglobin-analytics} shows the results of Judy's analytical inquiry into the three hemoglobin oxygen saturation hypotheses given the observations she managed to take from the literature. Unlike her expectations w.r.t. the principle of Occam's razor, Hill's model has been beaten by Dash's model, which is structurally more complex (viz., $|\mathcal S_{H}|=10$, $|\mathcal S_{D}|=35$). This top-ranked hypothesis includes additional observables such as \textsf{pCO$_2$} and \textsf{pH}. Hypothesis management can provide this and other model statistics (e.g., predictive power, mean computation time, etc), which may provide useful metrics to assess hypotheses qualitatively as well.

The $\Upsilon$-DB system can also be used to study a single hypothesis under very many alternative parameter settings (trials) aimed at finding the best fit for a phenomenon. For example, we have applied it for the case of the VPR's baroreflex hypothesis (Fig. \ref{fig:baroreflex}),\cite{beard2010} sized $|\mathcal S|=171$, to find the best fit among $1K$ trials stored in the probabilistic database.\cite[p. 81-2]{goncalves2015c} 

Altogether, it is worthwhile highlighting that $\Upsilon$-DB does not provide any new statistical tool for hypothesis testing. It rather can implement a suit of existing tools for enabling data-driven hypothesis management in a systematic fashion on top of state-of-the-art database technology.

\section{Prototype System}

\begin{figure}[t]
\advance\leftskip-1cm
\begin{subfigure}[t]{0.33\textwidth}
\includegraphics[width=0.96\textwidth]{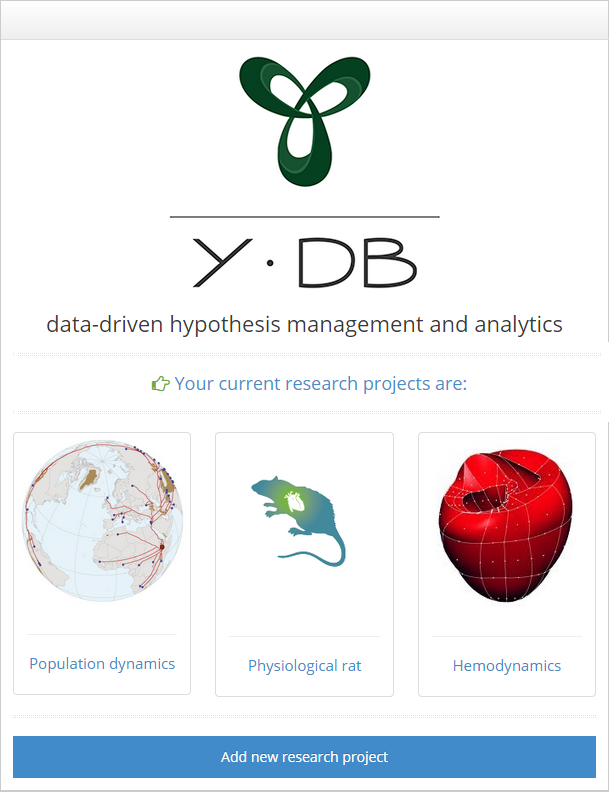}
\caption{Research dashboard.}
\label{fig:logo}
\end{subfigure}
\hspace{1pt}
\begin{subfigure}[t]{0.33\textwidth}
\includegraphics[width=1.09\textwidth]{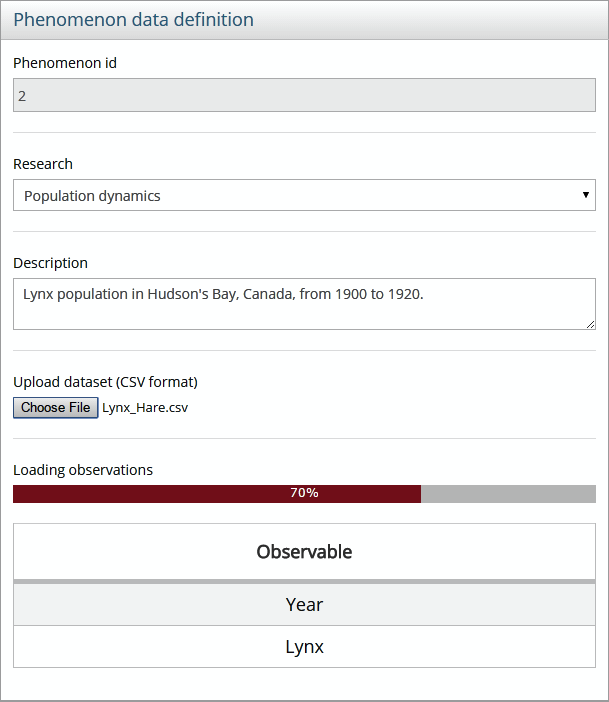}\\
\caption{Phenomenon data def.}
\label{fig:etl-phenomenon}
\end{subfigure}
\hspace{18pt}
\begin{subfigure}[t]{0.33\textwidth}
\includegraphics[width=1\textwidth]{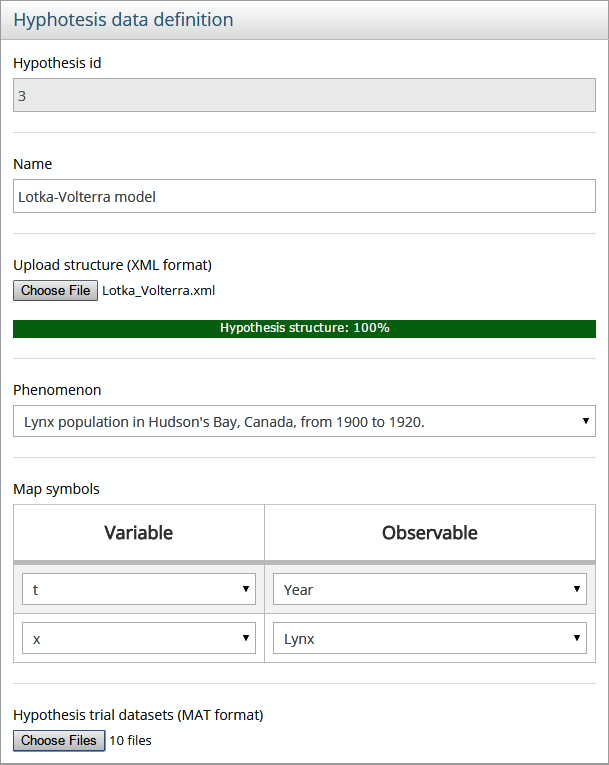}\\
\caption{Hypothesis data definition.}
\label{fig:etl-hypothesis}
\end{subfigure}\vspace{20pt}\\
\advance\leftskip 0.2cm
\begin{subfigure}[t]{0.33\textwidth}
\includegraphics[width=.91\textwidth]{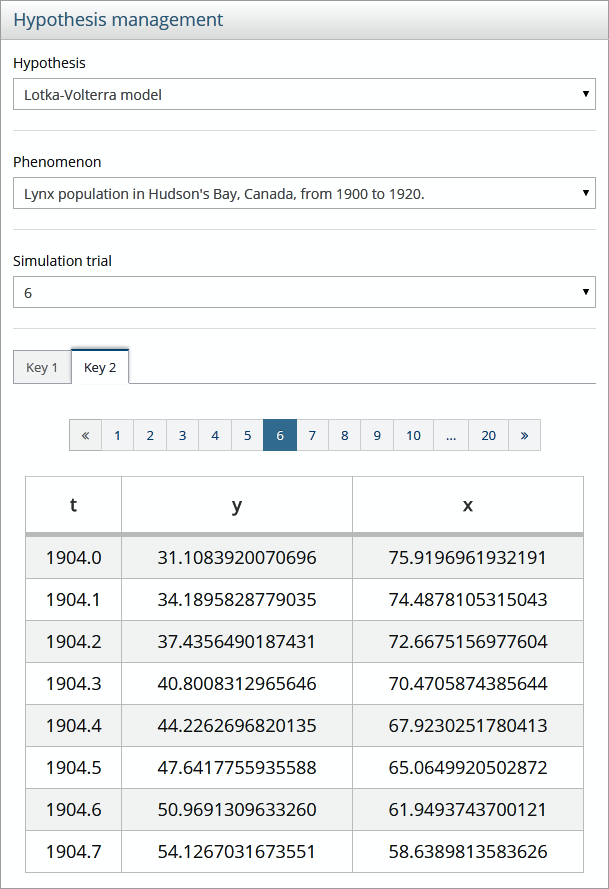}\\
\caption{Hypothesis data retrieval.}
\label{fig:management}
\end{subfigure}
\hspace{0pt}
\begin{subfigure}[t]{0.33\textwidth}
\includegraphics[width=1.05\textwidth]{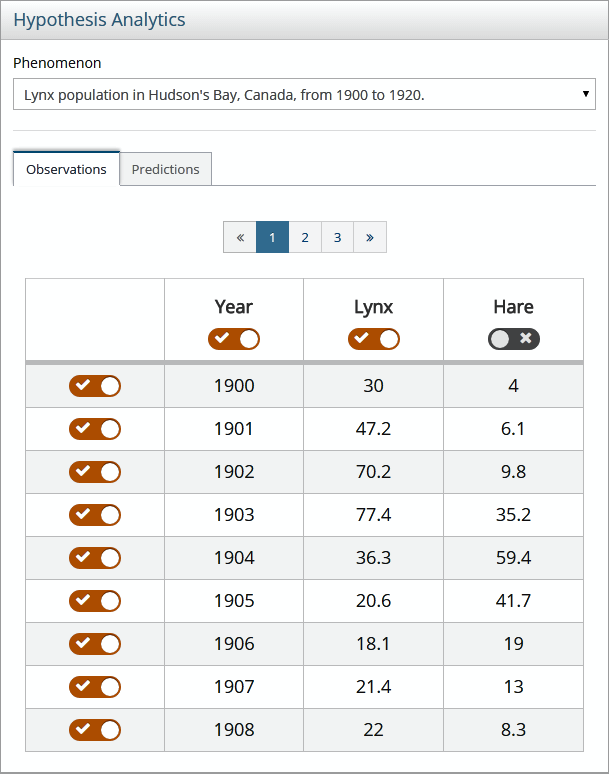}\\
\caption{Selected observations tab.}
\label{fig:analytics1}
\end{subfigure}
\hspace{13pt}
\begin{subfigure}[t]{0.33\textwidth}
\includegraphics[width=1.01\textwidth]{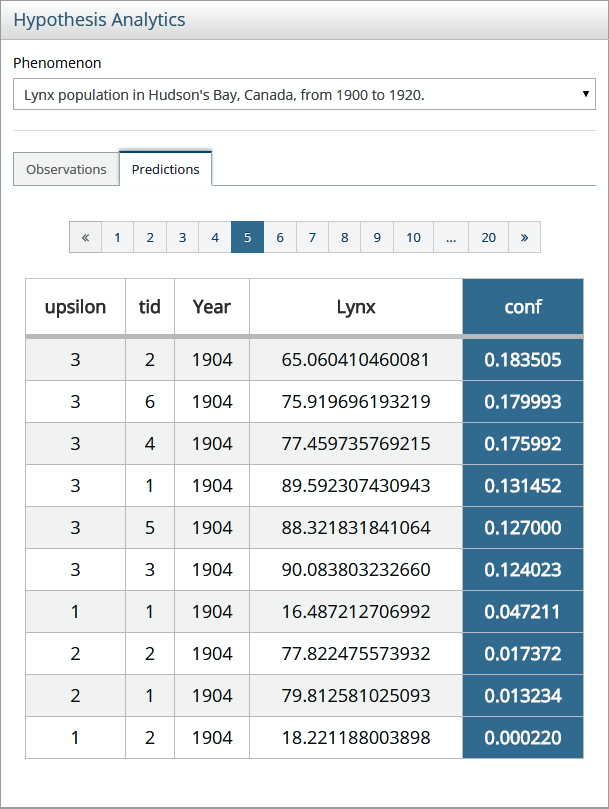}\\
\caption{Ranked predictions tab.}
\label{fig:analytics2}
\end{subfigure}
\caption{Screenshots of the $\Upsilon$-DB system prototype.}
\label{fig:demo}
\end{figure}

A first prototype of the $\Upsilon$-DB system has been implemented as a Java web application,\cite{goncalves2015b} with the pipeline component in the server side on top of \textsf{MayBMS} (a backend extension of \textsf{PostgreSQL}).\cite{koch2009} 
Fig. \ref{fig:demo} shows screenshots of the system in a population dynamics scenario comprising the Malthusian model, the logistic equation and the Lotka-Volterra model applied to predict the Lynx population in Hudson's Bay in Canada from 1900 to 1920. The observations, collected from Elton and Nicholson,\cite{elton1942} are used to rank the competing hypotheses and their trials accordingly.

Fig. \ref{fig:demo}(a) shows the research projects currently available for a user. Figs. \ref{fig:demo}(b, c) show the ETL interfaces for phenomenon and hypothesis data definition (by synthesis), and then the insertion of hypothesis simulation trial datasets. Note that it requires simple phenomena description, hypothesis naming and file upload to get phenomena and hypotheses available in the system to be managed as probabilistic data. 
Fig. \ref{fig:demo}(d) shows the interface for a basic retrieval of simulation data, given a selected phenomenon and a hypothesis trial. 
Figs. \ref{fig:demo}(e, f) show two tabs of the predictive analytics module. 
Note that the user chooses a phenomenon for study and imposes some selectivity criteria onto its observational sample. The system then lists in the next tab the corresponding predictions available, ranked by their probabilities conditioned on the selected observations. In this case, Lotka-Volterra's model (under trial $\textsf{tid}=2$) is the top-ranked hypothesis to explain the Lynx population observations in Hudson's Bay from 1900 to 1920.

\vspace{-9pt}
\section{Conclusions}
\vspace{-6pt}
Hypothesis data management is a promising new research field towards taking more value out of the theoretical data available in open simulation laboratories. 
Our work is a first effort to define its use case in the context of simulation data management. 
It proposes core principles and techniques for enconding and managing hypotheses as uncertain and probabilistic data,\cite{goncalves2014,goncalves2015c} enabling data-driven hypothesis testing and predictive analytics.\cite{goncalves2015b} 

A next step is to apply the method to hypotheses that are complex not only in terms of number of equations and coupled variables, but also dimensionally like in PDE models of fluid dynamics. Besides, major directions of future work are (i) to improve the statistical capabilities of the $\Upsilon$-DB system for supporting the data sampling out of simulation results, and (ii) to push its scalability forward to allow for hypothesis testing on samples of larger scale.


\section{Acknowledgments}
\vspace{-6pt}
This work has been supported by the Brazilian funding agencies CNPq (grants $\!$n$^o\!$ 141838/2011-6, 309494/2012-5) and FAPERJ (grants INCT-MACC E-26/170.030/2008, `Nota $\!$10' $\!$E-26/100.286/2013). $\!$We thank IBM for a Ph.D. Fellowship 2013-2014.

\bibliographystyle{abbrv}
\bibliography{cise}

\begin{thebibliography}{10}

\bibitem{ahmad2010}
Y.~Ahmad, R.~Burns, M.~Kazhdan, C.~Meneveau, A.~Szalay, and A.~Terzis.
\newblock Scientific data management at the {J}ohns {H}opkins {I}nstitute for
  {D}ata {I}ntensive {E}ngineering and {S}cience.
\newblock {\em {SIGMOD} Record}, 39(3):18--23, 2010.

\bibitem{ailamaki2010}
A.~Ailamaki, V.~Kantere, and D.~Dash.
\newblock Managing scientific data.
\newblock {\em Comm. ACM}, 53(6):68--78, 2010.

\bibitem{bassingthwaighte2000}
J.~B. Bassingthwaighte.
\newblock Strategies for the {P}hysiome {P}roject.
\newblock {\em Ann. Biomed. Eng.}, 28:1043--58, 2000.

\bibitem{bolstad2007}
W.~M. Bolstad.
\newblock {\em Introduction to Bayesian Statistics}.
\newblock Wiley-Interscience, 2nd edition, 2007.

\bibitem{beard2010}
S.~M. Bugenhagen, A.~W.~J. Cowley, and D.~A. Beard.
\newblock Identifying physiological origins of baroreflex dysfunction in
  salt-sensitive hypertension in the {Dahl SS} rat.
\newblock {\em Physiological Genomics}, 42:23--41, 2010.

\bibitem{cushing2013}
J.~B. Cushing.
\newblock Beyond big data?
\newblock {\em Computing in {S}cience \& {E}ngineering}, 15(5):4--5, 2013.

\bibitem{elton1942}
C.~Elton and M.~Nicholson.
\newblock The ten-year cycle in numbers of the lynx in {C}anada.
\newblock {\em Journal of Animal Ecology}, 11(2):215--44, 1942.

\bibitem{fregnac2014}
Y.~Fr\'egnac and G.~Laurent.
\newblock Where is the brain in the {H}uman {B}rain {P}roject?
\newblock {\em Nature}, 513:27--9, 2014.

\bibitem{goncalves2014}
B.~Goncalves and F.~Porto.
\newblock {$\Upsilon$-DB}: {M}anaging scientific hypotheses as uncertain data.
\newblock {\em PVLDB}, 7(11):959--62, 2014.

\bibitem{goncalves2015b}
B.~Goncalves, F.~C. Silva, and F.~Porto.
\newblock {$\Upsilon$-DB}: A system for data-driven hypothesis management and
  analytics.
\newblock Technical report, LNCC, 2015. (available at
  \href{http://arxiv.org/abs/1411.7419}{CoRR abs/1411.7419}).

\bibitem{goncalves2015c}
B.~Gon\c{c}alves.
\newblock {\em Managing large-scale scientific hypotheses as uncertain and
  probabilistic data}.
\newblock PhD thesis, National Laboratory for Scientific Computing (LNCC),
  Brazil, 2015. (available at \href{http://arxiv.org/abs/1501.05290}{CoRR
  abs/1501.05290}).

\bibitem{koch2009}
C.~Koch.
\newblock {\em May{BMS}: {A} system for managing large uncertain and
  probabilistic databases}.
\newblock In C. Aggarwal (ed.), Managing and Mining Uncertain Data, Chapter 6.
  Springer-Verlag, 2009.

\bibitem{markram2006}
H.~Markram.
\newblock The {B}lue {B}rain {P}roject.
\newblock {\em Nature Reviews Neuroscience}, 7:153--60, 2006.

\bibitem{meneveau2007}
E.~Perlman, R.~Burns, Y.~Li, and C.~Meneveau.
\newblock Data exploration of turbulence simulations using a database cluster.
\newblock In {\em Proc. of ACM/IEEE Supercomputing (SC'07)}, 2007.

\bibitem{simon1966}
H.~Simon and N.~Rescher.
\newblock Cause and counterfactual.
\newblock {\em Philosophy of Science}, 33(4):323--40, 1966.

\bibitem{ailamaki2013}
A.~Stougiannis, F.~Tauheed, M.~Pavlovic, T.~Heinis, and A.~Ailamaki.
\newblock Data-driven {N}euroscience: {E}nabling breakthroughs via innovative
  data management.
\newblock In {\em Proc. of the International Conference on Management of Data
  (SIGMOD '13)}, 2013.

\bibitem{suciu2011}
D.~Suciu, D.~Olteanu, C.~R\'e, and C.~Koch.
\newblock {\em Probabilistic Databases}.
\newblock Morgan \& Claypool Publishers, 2011.

\end{thebibliography}

\vspace{20pt}
\textbf{Bernardo Gon\c{c}alves} is currently a postdoctoral researcher at the University of Michigan, Ann Arbor. His research centers on the design of data systems and analytics to support the scientific method at scale. He holds a Ph.D. in Computational Modeling from the National Laboratory for Scientific Computing (LNCC) in Brazil, and a M.Sc. and a B.Sc. in Computer Science from the Federal University of Esp'rito Santo (UFES) in Brazil.

\vspace{20pt}
\textbf{Fabio Porto} is a researcher at LNCC, in Brazil. He holds a PhD in Computer Science from PUC-Rio, in a sandwich program at INRIA, France. Prior to LNCC, he worked as a researcher at EPFL, Database Laboratory, in Switzerland. At LNCC, he coordinates the Data Extreme Lab, whose focus is on  research and development  activities involving the design of techniques, algorithms and models for scientific data management and analysis. He is a member of the Brazilian Computer Society (SBC) and Association of Computing Machinery (ACM).

\end{document}